\documentclass[trackchanges,twocolumn]{aastex701}
\usepackage[caption = true]{subfig}
\usepackage{caption}
\usepackage{xcolor}
\renewcommand{\baselinestretch}{1.}

\newcommand{\solarmass}{M_{\odot}}

\newcommand{\mdot}{\dot{M}}

\begin{document}

\title{Why most neutron star low-mass X-ray binaries accrete transiently: an evolutionary study of transient and persistent phases}

\author[orcid=0009-0008-1113-0966]{Divyansh Tripathi}
\affiliation{Department of Physics, Indian Institute of Science Education and Research Bhopal, Bhopal, Madhya Pradesh, 462066, India}
\email[show]{tripathidivyansh2003@gmail.com}
\author[orcid=0000-0002-6351-5808]{Sudip Bhattacharyya}
\affiliation{Department of Astronomy and Astrophysics, Tata Institute of Fundamental Research, 1 Homi Bhabha Road, Mumbai, 400005, India}
\email{sudip@tifr.res.in}
\author[0009-0006-9192-4074]{Pulkit Ojha}
\affiliation{Center for Theoretical Physics, Aleja Lotnikow 32/46, Warsaw, 02-668, Poland}
\email{pojha@cft.edu.pl}

\begin{abstract}

A neutron star (NS) low-mass X-ray binary (LMXB), in which an NS accretes matter from a low-mass donor star, is an ideal source for probing some fundamental aspects of physics and astronomy, such as strong gravity, superdense matter, and the accretion-ejection processes.
However, to reliably achieve these goals, one must adequately understand NS LMXBs, including why some accrete persistently and others transiently.
Focused models, such as those based on a thermal-viscous instability in the accretion disk, are considered to explain transient accretion. However, broader perspectives, including which LMXB parameter values and phases cause transients and why there are more transients than persistents, remain poorly understood.
Here, our computation of the long-term evolution of NS LMXBs addresses these questions, providing insight into LMXB parameters and phases, naturally producing more transients than persistents, and being partially consistent with the known properties of observed sources.
For example, we typically find a greater fraction of persistent phase at lower orbital periods from the LMXB evolution computation, which is somewhat consistent with observations.
However, a lack of full consistency calls for improving the aforementioned focused models, and our computations provide a new way to discriminate among these models.

\end{abstract}

\keywords{\uat{Accretion}{14} --- \uat{Computational methods}{1965} --- \uat{High Energy astrophysics}{739} --- \uat{Low-mass x-ray binary stars}{939} --- \uat{Neutron stars}{1108}  --- \uat{Transient sources}{1851}}

\section{Introduction}\label{Sec: Intro}

A neutron star (NS) low-mass X-ray binary (LMXB) is a binary stellar system in which an NS accretes matter from a low-mass ($\lesssim 1.5\solarmass$) companion or donor star when it fills its Roche lobe \citep[RL; ][]{Bhattacharya-1991, Tauris2006book}. 
In such a system, the NS spins up to millisecond periods due to angular momentum transfer as it accretes matter from the donor star via an accretion disk. 

Sometimes, these NSs appear as accreting millisecond X-ray pulsars (AMXPs), as X-rays are produced when the accretion disk is truncated by the NS magnetosphere and the material is channeled along the magnetic field lines of the star  
\citep{disalvo2020accretionpoweredxraymillisecond,Patruno2020}.
Some of the NS LMXBs remain in 
a quiescent state most of the time (for $\sim$~years) and undergo an outburst (for $\sim$~days to months) when the accretion onto the NS increases by orders of magnitude, possibly due to thermal-viscous instabilities in the accretion disk \citep[e.g., ][and references therein]{Done-2007A&AR,Bhattacharya:2025qps}. 
Such sources are known as transient LMXBs. 
According to the widely accepted thermal-viscous instability model, a NS accretes transiently through cycles of outbursts and quiescent periods, when the long-term average accretion rate ($\dot M_{\rm av}$) is below a critical rate $\dot M_{\rm av,crit}$ \citep[e.g., ][]{Dubus-1999, Lasota_1997, King-1996ApJ, 2021MNRAS.502L..45B}.
But for $\dot M_{\rm av} > \dot M_{\rm av,crit}$, the NS accretes in a relatively steady manner, and these are persistent NS LMXBs. 
At the end of the LMXB phase, many NSs become radio millisecond pulsars \citep[RMSPs; ][]{BhaswatiBhattacharyya_Jayanta_Roy_2021_Radio_Millisecond_pulsars,Tauris_Langer-2012MNRAS}.

The detection of NS LMXBs and their population studies depend on whether sources accrete transiently or persistently. 
This is because, typically, transient sources are detected, or at least identified as NS LMXBs, when they are sufficiently bright during an outburst.
Moreover, while one can study accretion processes in the strong-gravity regime and probe related parameter values over a relatively narrow range of the instantaneous accretion rate ($\dot{M}$) for persistent sources, transients allow such a study over a large $\dot{M}$ range, thus providing a broader understanding. However, one must understand the origin of the transient phenomena to interpret such a study correctly. 
While the aforementioned thermal-viscous instability model may explain these phenomena in many sources, it may not apply to some transient NS LMXBs \citep{Bhattacharya:2025qps}.
Furthermore, there are different proposed expressions of $\dot M_{\rm av,crit}$ for the thermal-viscous instability model \citep[e.g., ][]{Dubus-1999, Lasota_1997, King-1996ApJ, 2021MNRAS.502L..45B}, and they need to be observationally constrained.

In addition, merely modeling the currently observed transient and persistent accretion properties yields an incomplete understanding of NS LMXBs, with several questions left poorly explored. 
For example, why are there more transient NS LMXBs and fewer persistent ones? 
Which source parameters and phases of the LMXB states give rise to transients, and which ones give rise to persistents?
Without answers to these questions, one remains confined to relatively narrow or focused aspects of NS LMXBs and misses the wider perspective.
Thus, to gain deeper and wider insight, one needs to explore which donor stars, initial conditions, and other source parameters can produce transient and persistent accretion across different phases of the LMXB state, and to compare these numerical results with the properties of the known sources.

Such exploration is essential also for the following reasons.
It was shown that the evolution of the NS spin frequency ($\nu$) and other properties crucially depends on whether the accretion is transient or persistent \citep{Sudip-2017ApJ,2021MNRAS.502L..45B}.
Therefore, in order to explain the distribution of $\nu$ and other NS and binary parameter values among RMSPs and NS LMXBs to probe the fundamental physics of NSs and plausible emission of continuous gravitational waves due to the mass quadrupole moment of spinning NSs \citep{Bhattacharyya2010,BhattacharyyaJ1023,Bhattacharyya2017AQX-1}, one needs to compute the evolution of the binary system and the NS throughout the LMXB state and explore the nature of accretion, persistent or transient, in various phases.

Several previous papers have studied the evolution of the binary and the donor star parameters \citep{2013ApJ...775...27C,2015ApJ...814...74J,He-2019RXiang}, and also of the NS parameters \citep{2024MNRAS.535..344K,kar2025longtermevolutionscox1}

throughout the LMXB state.
However, in this work, for the first time to the best of our knowledge, we systematically explore the phases of persistent accretion and transient accretion throughout the LMXB state and compare our results with a sample of known NS LMXBs.

We provide a brief discussion of our methods for calculating LMXB evolution in section~\ref{sec: Methods} and present the results in section~\ref{sec: Results}. In section \ref{Sec: Discussion and Conclusion}, we discuss our results and conclude. 

\section{Formalism} \label{sec: Methods}

We calculate the evolution of LMXBs using MESA-r23.05.1 \citep{paxton_2023_7983526}\citep[Modules for Experiments in Stellar Astrophysics; ][]{Paxton2011, Paxton2013, Paxton2015, Paxton2018, Paxton2019, Jermyn2023}, which is a 1D stellar evolution code.
The binary system consists of an NS (mass: $M_1$), considered to be a point mass, and a donor star (mass: $M_2$), initially in the zero-age main sequence (ZAMS) with the solar composition \citep{2015ApJ...814...74J}. 
The computation of binary evolution was described in detail in a previous paper \citep{2024MNRAS.535..344K}, which we briefly mention here.
The input parameter values are given in Table~\ref{tab: Parameter_Table}. 
We compute the binary evolution for a grid of initial values of the orbital period ($P_{\rm orb}$) and $M_2$ (see Table~\ref{tab: Parameter_Table} for ranges of initial values).

\begin{table}[!h]
    \caption{A list of input parameters used for the computation with MESA, and their used values or ranges.}
    \begin{tabular}{l l}
    \hline
     Parameter &  Values\\\hline
    Initial orbital period $P_{\text{orb}}$ (day) & 0.5-2.5\\
    Initial donor mass $M_{2}$ $(M_\odot)$ & 
    0.5--1.5\\
   Initial NS mass $M_1$ $(M_\odot)$ & 1.35\\
    Magnetic braking Index ($\gamma$) & 4 \\
    Fractional mass-loss ($\beta$) & 0.5 \\ \hline
    \end{tabular}
    \label{tab: Parameter_Table}
\end{table}

The donor star Roche lobe radius is given by \citet{1983ApJ...268..368E} as 
\begin{equation}\label{eq:1}
    \frac{R_{\rm{L,2}}}{a} = \frac{0.49q^{\frac{-2}{3}}}{0.6q^{\frac{-2}{3}} + \ln(1+q^{\frac{-1}{3}})},
\end{equation}
where $q= \frac{M_1}{M_2}$, and $a$ is the orbital separation. Mass transfer through the Roche lobe overflow is calculated using an expression from \citet{1983ApJ...268..368E},
\begin{equation}\label{eq:2}
    \mdot_2 = \mdot_{0} \exp \left[\frac{R_2 - R_{\rm{L,2}}}{H_{\rm{p}}{/\gamma}(q)}\right],
\end{equation}
where $H_{\rm p}$ is the pressure scale height of the atmosphere of the donor star and $\gamma$ is a function of $q$. $\mdot_2$ and $R_2$ are the donor star's mass loss rate and radius, respectively. 
Here, we consider three commonly used mechanisms of the orbital angular momentum ($J$) loss rate (AML): gravitational radiation (GR) due to orbital motion, mass loss (ML) from the system, and magnetic braking (MB).
Therefore,
\begin{equation}\label{eq:3}
    \dot{J} = \dot{J}_{\rm{GR}} + \dot{J}_{\rm{ML}} + \dot{J}_{\rm{MB}}.
\end{equation}
Here, $\dot{J}_{\rm{GR}}$ is given as \citep{Landau:1975pou}
\begin{equation}\label{eq:4}
    \dot{J}_{\rm{GR}} = -\frac{32}{5}\frac{G^{7/2}}{c^5}\frac{M_1^2 M_2^2 (M_1 + M_2^{1/2})}{a^{7/2}},
\end{equation}
where, $G$ and $c$ are the gravitational constant and the speed of light in vacuum, respectively. 
The AML due to magnetic braking is given by \citep{1983ApJ...268..368E}
\begin{equation}\label{eq:5}
    \dot{J}_{\rm{MB}} = -3.8 \times 10^{30} M_2 R_2^{\gamma} \omega^3 \textrm{dyn} . \textrm{cm},
\end{equation}
where we set the magnetic braking index $\gamma = 4$, \citep{2013ApJ...775...27C, 2015ApJ...814...74J}. Here, $\omega$ is the spin frequency of the donor star. 
We also assume that the MB switches off when the star's convective envelope becomes too thin and starts when the convective core mass fraction is less than $ 0.02$.  
The AML due to mass loss is given as 
\begin{equation}\label{eq:6}
    \dot{J}_{\rm{ML}} = -\beta \dot{M}_2\left(\frac{M_2}{M_1 + M_2}\right)^2 a^2 \omega,
\end{equation}
where $\beta$, which is set to 0.5 for this work, is the fraction of mass lost from the proximity of the accretor due to winds. 
Hence, the long-term average accretion rate on the NS ($\dot{M}_{\rm av}$) can be given as 
\begin{equation}
    \dot{M}_{\rm av} = \beta \dot{M}_2.
\end{equation}

In our model, $\dot{M}$ is limited by the Eddington limit. \citep[e.g., ][]{2015ApJ...814...74J}
\begin{eqnarray}\label{eq:7}
   \dot{M}_{\rm{Edd}} = 3.6\times10^{-8}\left(\frac{M_1}{1.4\solarmass}\right)
    \left(\frac{0.1}{GM_1/R_1c^2}\right) \\ \nonumber
    \left(\frac{1.7}{1+X}\right)\solarmass {\rm, yr}^{-1},
\end{eqnarray}
where $R_1$ is the NS radius, and $X$ is the hydrogen abundance.
In our computation, we include irradiation of the donor star due to X-ray luminosity, which is given as 
\begin{equation}\label{eq:10}
    L_X = \frac{GM_1\dot{M}}{R_1}.
\end{equation}
The corresponding flux for irradiation is considered to be  

\begin{equation}\label{eq:11}
    F_{\rm{irr}} = \epsilon\frac{L_X}{4\pi a^2},
\end{equation}
where $\epsilon$ is the irradiation efficiency. 
Motivated by \citet{Goodwin-2020MNRAS}, we limit the maximum flux of irradiation to $3.1\times10^9~\textrm{erg s$^{-1}$ cm$^{-2}$}$ and the irradiation distance to $10^{13}\textrm{cm}$.

Moreover, while accretion rate is the baryonic mass-transfer rate, the gravitational mass of the NS should be considered for the computation of binary evolution. We take care of this following \cite{2024MNRAS.535..344K}, using the conversion between the baryonic mass and the gravitational mass given in \citet{2015PhRvD..92b3007C}.

In order to determine when accretion onto the NS is transient and when it is persistent, we need to use an expression of $\dot M_{\rm av,crit}$. 
As mentioned in section~\ref{Sec: Intro}, various expressions of $\dot M_{\rm av,crit}$ have been proposed in the literature depending on the model.
We adopt two standard formulations for the critical average mass accretion rate ($\dot{M}_{\rm av, crit}$) derived from the thermal-viscous disk instability model (DIM) for irradiated accretion disks, to test the distribution of transient and persistent accretion phases throughout the low-mass X-ray binary (LMXB) population.

The first formulation is obtained by comparing the effective irradiation temperature ($T_{\rm irr}$) at the disk outer edge to the hydrogen ionization temperature ($T_{\rm H}$), where $T_{\rm H}\sim6500$ K. 
The disk is considered to be stable if $T_{\rm irr} > T_{\rm H}$ for the entire disk.
This criterion gives the critical average mass accretion
rate \citep{Paradijs-1996ApJ, King-1996ApJ, Lasota_1997}:

\begin{equation}\label{critical_1}
\dot{M}_{\rm{av, crit1}} = 5\times 10^{-11} \left( \frac{M_1}{M_\odot} \right)^{\frac{2}{3}} \left( \frac{P_{\rm{orb}}}{3\text{ hr}} \right)^{\frac{4}{3}} M_\odot\text{ yr}^{-1}.
    \end{equation}

The second formulation \citep{Dubus-1999} is similar to the above, but here the authors self-consistently consider also the effects of the irradiation on the disk, for example, on its height at a given radius.  
This calculation, for specific parameter values \citep[e.g., $C = 5\times10^{-4}$ in Eq.~32 of][]{Dubus-1999}, gives the critical average mass accretion rate:

\begin{eqnarray}\label{critical_2}
\dot{M}_{\rm{av, crit2}} = 3\times 10^{-9} \left( \frac{M_1}{1.4\,M_\odot} \right)^{0.5} \left( \frac{M_2}{\,M_\odot} \right)^{-0.2} \\ \nonumber
\left(\frac{P_{\rm{orb}}}{1.0\text{ d}} \right)^{1.4} 
M_\odot\text{ yr}^{-1}.
    \end{eqnarray}

Throughout the computation of binary evolution, we compare $\dot{M}_{\rm{av, crit1}}$ and $\dot{M}_{\rm{av, crit2}}$ with $\dot{M}_{\rm{av}}$ to determine if the source is in transient or persistent phase for each expression (Eqs.~\ref{critical_1} or \ref{critical_2}) of the critical accretion rate.
We compare our results with the $M_2$ and $P_{\rm orb}$ values of observed NS LMXBs.
For this, we select persistent and transient sources (Table~\ref{Table1: Transient LMXB}) with known $M_2$ and $P_{\rm orb}$ values.

\startlongtable
\begin{deluxetable*}{ccccc}
\tablecaption{A sample of transient and persistent NS LMXBs. \label{Table1: Transient LMXB}}
\tablewidth{0 pt}
\tablehead{
\colhead{S.No.} &
\colhead{Name} &
\colhead{$M_2$ ($M_\odot$)} &
\colhead{$P_{\rm orb}$ (d)} &
\colhead{Reference}
}
\startdata
\multicolumn{5}{c}{Transient LMXBs}\\
\tableline
1&IGR-J17480-2446&0.45&0.9&1,2,3 \\
2&Swift J1858.6-0814&0.56&0.89&2\\
3&Aq1-X1&0.5&0.78951&1,2,4\\
4&Cen X-4&0.3&0.63&2\\
5&4U 1608-52&0.32&0.54&2\\
6&GRS 1747-312&0.1-0.8&0.515&2 \\
7&IGR J18245-2452&0.2&0.4594& 1,2 \\
8&SWIFT J1749.4-2807&1.0&0.37&2\\
9&4FGL J0427.8-6704&0.3&0.37&2\\
10&Swift J1749.4-2807&0.46-0.81&0.36736&1,2 \\ 
11&SAX J1748.9-2021&0.70-0.83&0.365&1,2 \\ 
12&IGR J17591–2342&0.42&0.36&1,2 \\ 
13&AX J1745.6-2901&0.8&0.35&2\\
14&MXB 1659-29&0.9&0.3&2\\
15&XSS J12270-4859&0.46&0.29&2\\
16&XTE J2123-058&0.6&0.25&2\\
17\large{$^*$}&4U 2129+47&0.4-0.8&0.22&6\\ 
18&XTE J1719-291&0.6&0.21&2\\
19&MAXI 1J1816-195&0.1-0.55&0.201&2,3 \\ 
20&PSR J1023+0038&0.2&0.2&2\\
21&XTE J1814-338&0.19-0.32&0.178&1,2 \\ 
22&IGR J17498–2921&0.17-0.48&0.160134&1,2\\ 
23\large{$^*$}&EXO 0748-676&0.62&0.16&2\\ 
24&IGR J17511-3057&0.13&0.14453&1,2 \\ 
25&IGR J00291+5934&0.039-0.0995&0.102083&1,2 \\ 
26&GS 1826-238&0.2&0.09&2\\
27&SAX J1808.4-3658&0.05&0.0839&1,2 \\ 
28&IGR J17379–3747&0.06&0.0791666&1,2 \\ 
29&4U 1905+000&0.05&0.06&2\\
30\large{$^*$}&HETE J1900.12455&0.016-0.07&0.057&1,2 \\ 
31\large{$^*$}&1H 1905+000&0.05&0.0554&2\\ 
32&IGR J17494-3030&0.02&0.0520&2,3 \\ 
33&4U 1905+000&0.05&0.05&2,3\\
34\large{$^*$}&1M 1716–315&0.01&0.04&7\\ 
35&MAXI J1957+032&0.043-0.085&0.04&1,2 \\ 
36&NGC6440 X–2&0.0067&0.03958&1,2 \\ 
37&Swift J1756.9–2508&0.006-0.03&0.0379&1,2 \\ 
38&IGR J16597-3704&0.0065&0.0319&1,2 \\ 
39&MAXI J0911-655&0.024&0.0307&1,2 \\ 
40&XTE J0929-314&0.0083&0.0304&1,2 \\ 
41&XTE J1751–305&0.014-0.058&0.029&1,2 \\ 
42&XTE J1807-294&0.006&0.027&1,2\\ 
43\large{$^*$}&IGR J17062-6143&0.006-0.0216&0.02636&1,2\\ 
\tableline
\multicolumn{5}{c}{Persistent LMXBs}\\
\tableline
1&Cyg X-2 & 0.5799 & 9.84166 & 5 \\ 
2&1RXH J173523.7-354013 & 1.4 & 7.5 & 5 \\
3&4U 1705-44&0.5&0.41&2,5\\
4&2A 1822-371 & 0.5 & 0.33 & 5,1,4 \\ 
5&4U 1254-69 & 0.42 & 0.16375 & 5 \\ 
6&4U 1636-536 & 0.36 & 0.15833 & 4\\ 
7&4U 1323-62 & 0.28 & 0.12208 & 4 \\ 
8&XB 1832-330 & 0.026 & 0.089583 & 5 \\ 
9&Ser X-1 & 0.14 & 0.0833 & 5\\
10&4U 0614+09 & 0.0142 & 0.035635 & 5 \\ 
11&4U 1915-05&0.0144&0.03458&2,5\\
12&4U 1626-67 & 0.04 & 0.028749 & 4 \\ 
13&M15 X-2 & 0.03 & 0.01566 & 5 \\ 
14&4U 1850-087 & 0.0356 & 0.01429 &5  \\ 
15&4U 1543-624 & 0.0405 & 0.01262 & 4 \\ 
16&2S 0918-549 & 0.024-0.039 & 0.0120 & 5 \\ 
17&4U 0513-40 & 0.045 & 0.0118 & 5 \\ 
18&4U 1820-303 & 0.0651 & 0.00791 & 5 \\
\enddata
\tablecomments{
Entries marked with $^{*}$ corresponds to long-outburst systems \citep{Bhattacharya:2025qps}.
}
\tablerefs{1. \citet{disalvo2020accretionpoweredxraymillisecond},
2. \citet{2023AandA...675A.199A},
3. \citet{heinke2024catalogoutburstsneutronstar},
4. \citet{2007AandA...469..807L},
5. \citet{2020ApJS..249...32G},
6. \citet{2008AABothwell},
7. \citet{jonker:hal-03801085}}
\end{deluxetable*}

\begin{figure*}[ht]
\centering 
\subfloat[]{
  \raggedleft{\includegraphics[width=83mm]{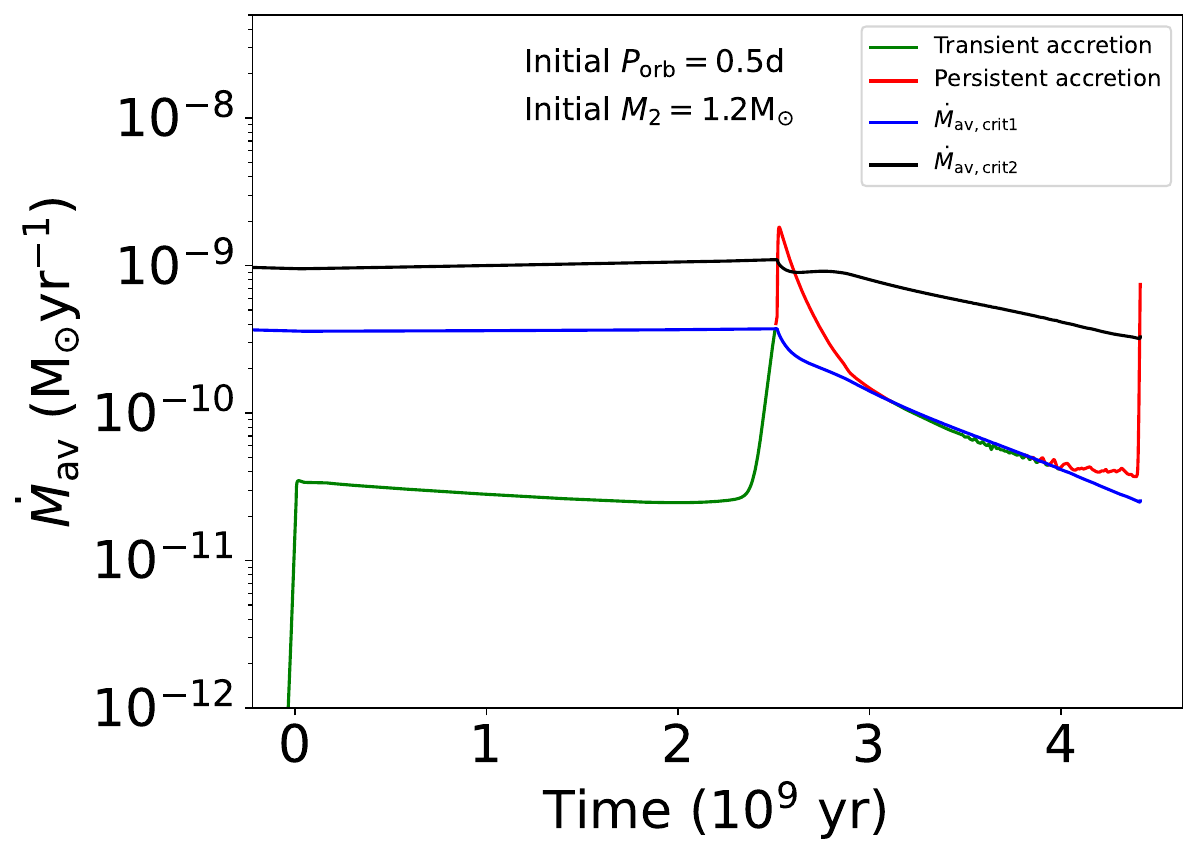}}
  \label{fig: P=0.5}
}\hfill
\subfloat[]{
  \includegraphics[width=80mm]{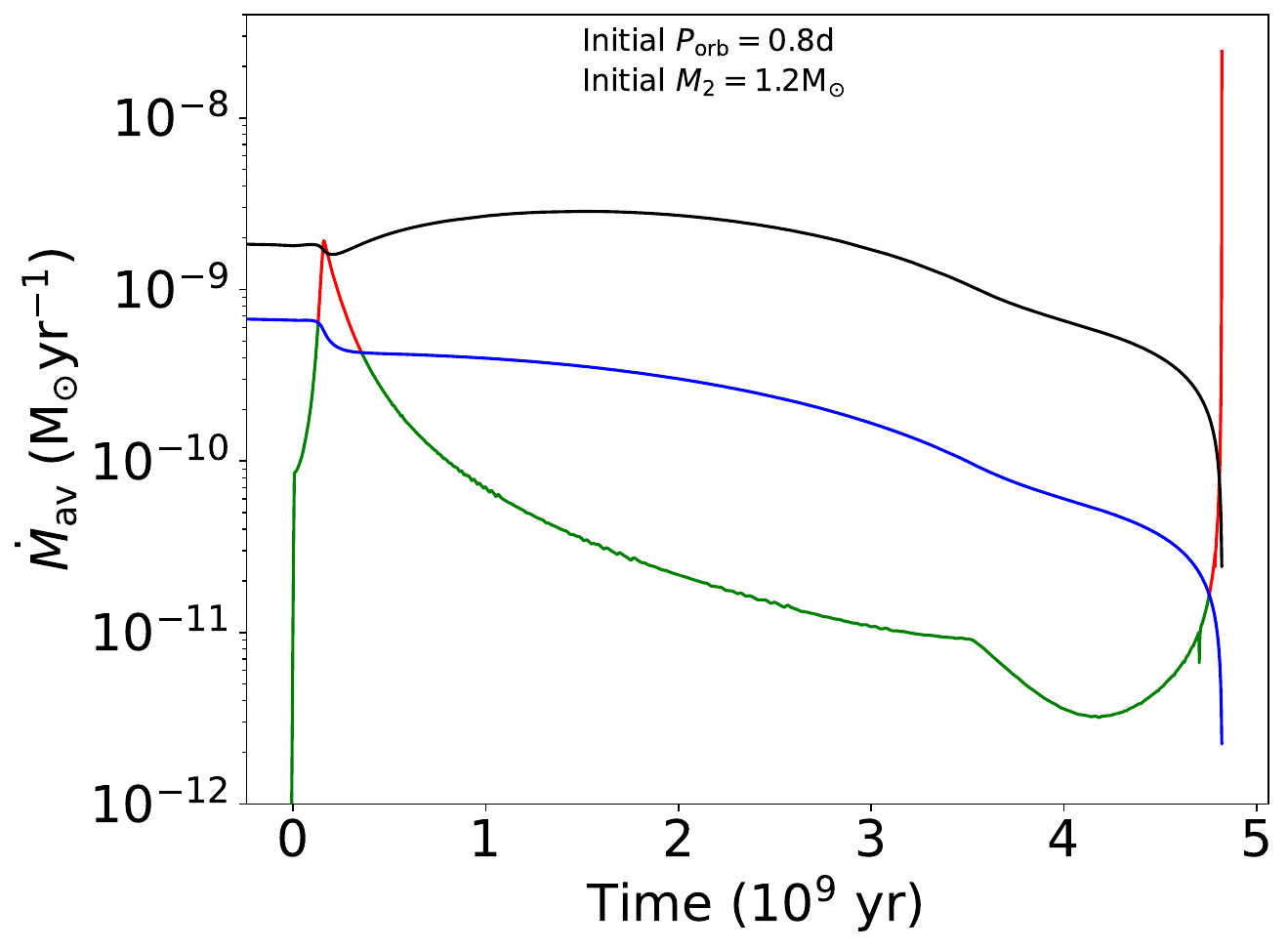}
  \label{fig: P=0.8}
}\\
\subfloat[]{
\vspace{2mm}
\includegraphics[width=80mm]{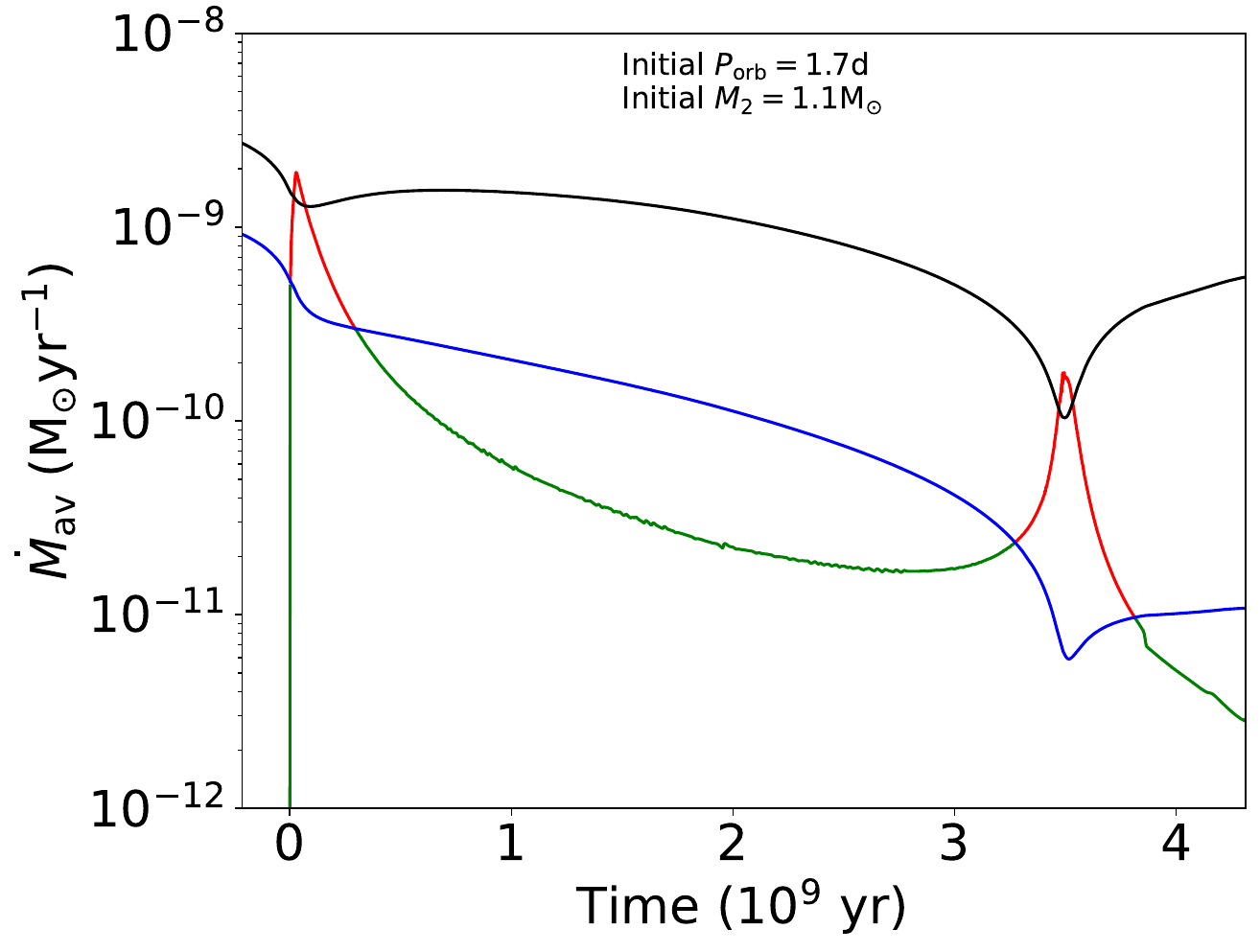}
\label{fig: P=1.7}
}\hfill
\subfloat[]{
\includegraphics[width=80mm]{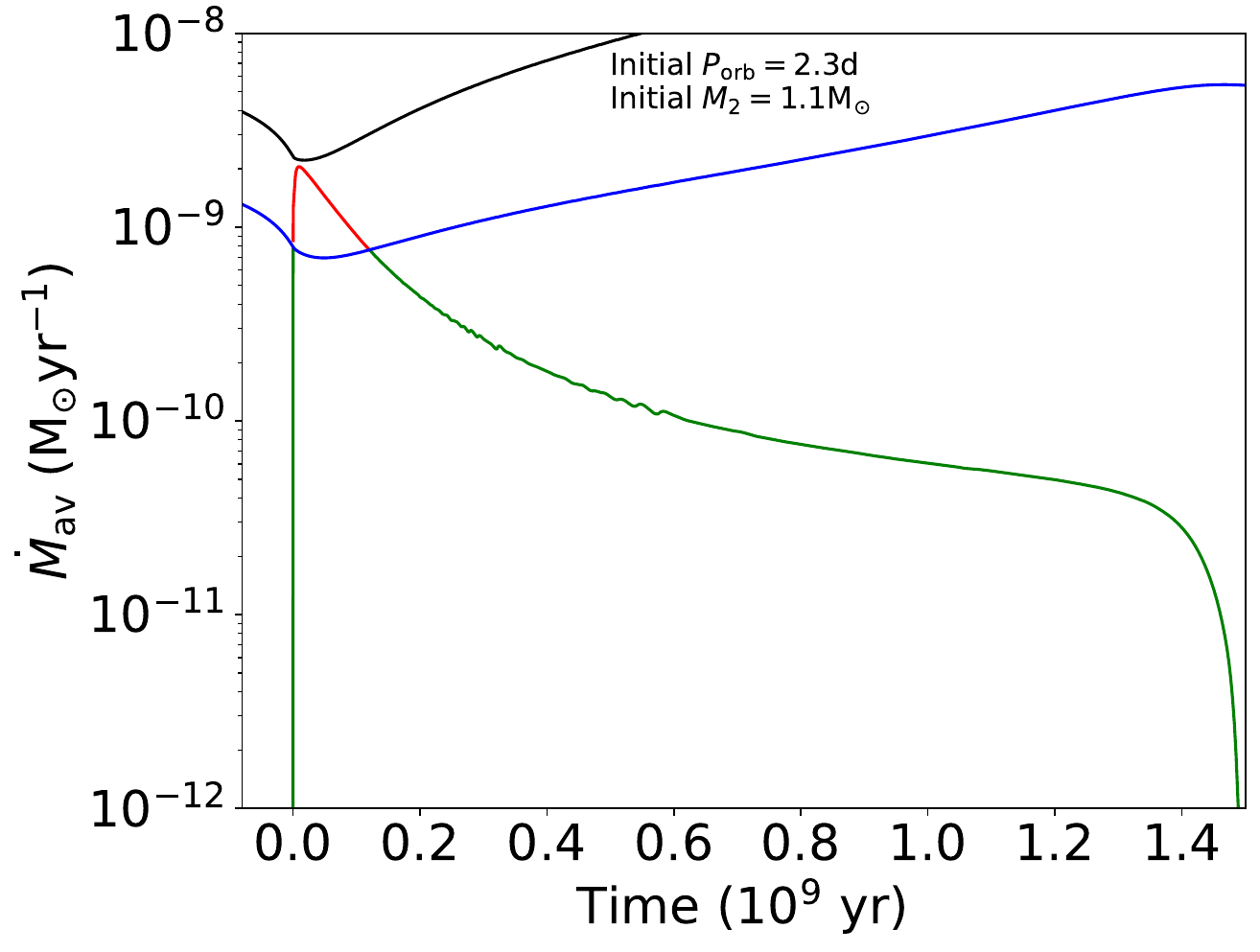}
\label{fig: P=2.3}
}
\caption{Computation of time evolution of the long-term average accretion rate ($\dot {M}_{\rm av}$) in neutron star low-mass X-ray binaries using MESA (see section \ref{sec: Results}). The four panels are for the initial values of $M_2$ and $P_{\rm{orb}}$ mentioned on the panels. 
In each panel, time starts at the onset of accretion. 
Blue ($\dot{M}_{\rm av,crit1}$; Eq. \ref {critical_1}) and black ($\dot{ M}_{\rm av,crit2}$; Eq. \ref {critical_2}) curves denote the evolution of the critical accretion rates. 
The $\dot{M}_{\rm av}$ evolution curve is green for transient accretion (considering $\dot{M}_{\rm av} < \dot{M}_{\rm av,crit1}$) and red for persistent accretion (considering $\dot{M}_{\rm av} > \dot{M}_{\rm av,crit1}$). 
This figure shows the phases of transient and persistent accretion across different parameter values, assuming Eq.~\ref{critical_1}.}
\label{fig: Accretion Curve}
\end{figure*}

\section{Results}\label{sec: Results}

\begin{figure*}[!h]
	\centering
\gridline{\fig{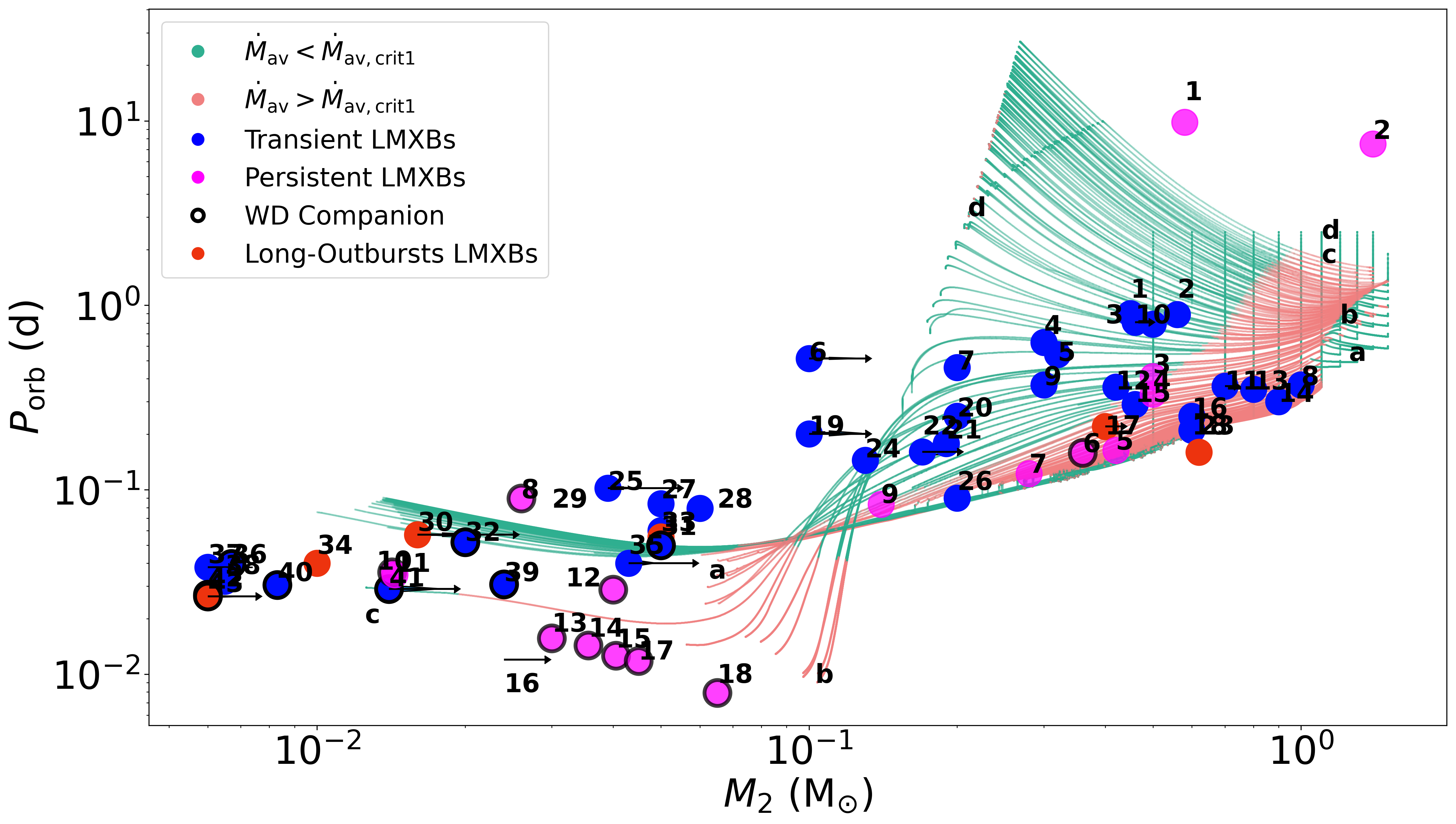}{0.85\linewidth}{(a)}}
\gridline{\fig{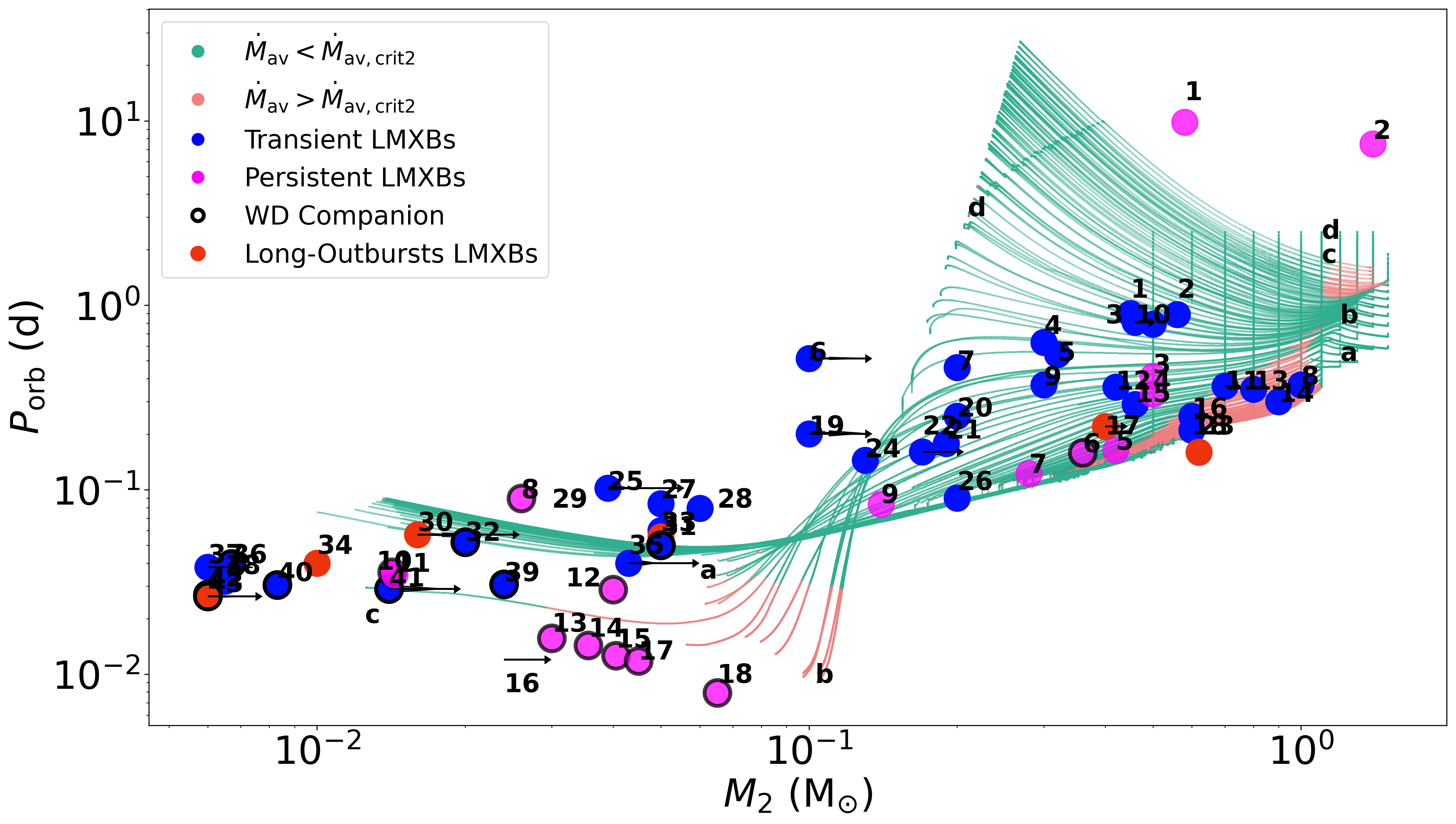}{0.85\linewidth}{(b)}}
\caption{Orbital period versus donor mass plot from the computation of time evolution of neutron star (NS) low-mass X-ray binaries (LMXBs)
using MESA (see section \ref{sec: Results}). 
The upper and lower panels are for $\dot{M}_{\rm av,crit1}$ (Eq.~\ref{critical_1}) and $\dot{M}_{\rm av,crit2}$ (Eq.~\ref{critical_2}), respectively.
In each panel, the green and light-red portions of the curves indicate theoretically predicted transient and persistent accretion, respectively. 
For a comparison of these theoretical curves with observed sources, known transient (blue-filled circle) and persistent (pink-filled circle) NS LMXBs with measured/estimated orbital period and donor mass values (or limits) are placed in the panels. Here, the black boundary implies a white dwarf companion, and red-filled circles denote sources showing long outbursts. The observed sources are identified by number (compare with Table~\ref{Table1: Transient LMXB}).
The letters a, b, c, and d on either side of the curves correspond to panel numbers of Fig.~\ref{fig: Accretion Curve} and denote the corresponding curve for the initial orbital period and donor mass.}

\label{fig: Orbital_Donor Mass}
\end{figure*}

As mentioned in section~\ref{sec: Methods}, we compute the binary evolution using MESA for a grid of initial values of $P_{\rm orb}$ and $M_2$. 
In Fig.~\ref{fig: Accretion Curve}, we show examples of $\dot M_{\rm av}$, $\dot M_{\rm av,crit1}$, and $\dot M_{\rm av,crit2}$ evolution during the LMXB state for four different combinations of initial values of ($P_{\rm orb}, M_2$).
In Fig.~\ref{fig: Orbital_Donor Mass}, we present all the computations of binary evolution in the $P_{\rm orb}$ versus $M_2$ plane. Here, the evolutionary curves and the end products differ significantly, depending on the initial ($P_{\rm orb}$, $M_2$) values. The four panels of Fig.~\ref{fig: Accretion Curve} represent four such different categories.

Fig.~\ref{fig: P=0.5} shows the evolution of an NS LMXB with initial $M_2 = 1.2~M_\odot$ and initial $P_{\rm orb} = 0.5$~d, and this evolution is marked as ``a'' in Fig.~\ref{fig: Orbital_Donor Mass}.  
As the Roche lobe radius equals the donor star radius, the accretion begins. 
Initially, the NS accretes transiently, followed by a brief period of persistent accretion driven by magnetic braking.
The $\dot M_{\rm av}$ increases again towards the end of the accretion phase due to the increase of $|\dot{J}_{\rm{GR}}|$ and the resulting rapid decrease of the binary separation.
The system remains transient for about 86\% of the total time of the LMXB state, considering Eq.~\ref{critical_1}, while this time is around 97\% if we consider Eq. \ref{critical_2}. 

Fig.~\ref{fig: P=0.8} shows the evolution of an NS LXMB with an initial donor mass of 1.2~$M_\odot$ and an initial orbital period of 0.8~d (marked as ``b'' in Fig.~\ref{fig: Orbital_Donor Mass}). 
The $\dot M_{\rm av}$ increases again towards the end due to a reason similar to that for Fig.~\ref{fig: P=0.5}.
Considering the critical rate of Eq.~\ref{critical_1}, the system remains transient for 95\% of the total time.
For the critical rate of Eq.~\ref{critical_2}, the source is transient for almost the entire period of the LMXB state.

Fig. \ref{fig: P=1.7} shows accretion curves for an NS LXMB with an initial donor mass of 1.1~$M_\odot$ and an initial period of 1.7~d (marked with ``c" in Fig.~\ref{fig: Orbital_Donor Mass}).  
Here, both critical accretion rates give two periods of persistent accretion. According to Eq.~\ref{critical_1}, transient accretion occurs for about 88\% of the LMXB state, while it is 97\% if one considers Eq.~\ref{critical_2}. 
Such systems evolve to very low orbital periods (less than an hour) and very low donor masses ($\sim 0.01 M_\odot$).

Fig. \ref{fig: P=2.3} shows accretion curves for an NS LMXB with initial donor mass and period of 
1.1~$M_\odot$ and 2.3~d, respectively (marked with ``d" in Fig.~\ref{fig: Orbital_Donor Mass}). 
. 
From the accretion rate curves, we see transient accretion for about 94\% of the LMXB state, assuming Eq.~\ref{critical_1}, whereas always a transient phase for Eq.~\ref{critical_2}. 

Fig.~\ref{fig: Orbital_Donor Mass} displays a grid of evolutionary tracks showing the persistent (red) and transient (green) phases of accretion. 
The tracks show the effect of the bifurcation period \citep[$P_{\rm bif}$; ][]{Bhattacharya-1991} with the curves for $P_{\rm orb} > P_{\rm bif}$ going upwards, i.e., towards longer periods, and those for $P_{\rm orb} < P_{\rm bif}$ going downwards, i.e., towards shorter periods. 

In Fig.~\ref{fig: Orbital_Donor Mass}, a comparison between the calculated accretion rate and the critical accretion rate (Eqs.~\ref{critical_1} and \ref{critical_2} for upper and lower panels, respectively) determines the transient and persistent phases. 
For both critical accretion rate models, the common trend for NS LMXBs is as follows (see Fig.~\ref{fig: Orbital_Donor Mass}). (1) Overall, the sources spend much more time in the transient phase than in the persistent phase.
(2) At the onset of the LMXB phase, $\dot M_{\rm av}$ is low, and hence the source is transient. (3) As $\dot M_{\rm av}$ rises, the source typically accretes persistently. (4) However, with the decline of $\dot M_{\rm av}$ at a later time, the source again becomes a transient.
(5) The persistent phase lasts longer in sources with low initial $P_{\rm orb}$ that evolve to even lower $P_{\rm orb}$. The last aspect is due to the fact that the critical accretion rate is lower for lower $P_{\rm orb}$ values (Eqs.~\ref{critical_1} and \ref{critical_2}), and hence the chances that the condition for persistent accretion, viz., $\dot M_{\rm av} > \dot M_{\rm av,crit}$, is satisfied are greater.
Moreover, the contribution of gravitational radiation to the orbital angular momentum loss rate is significant for lower $P_{\rm orb}$ values \citep{Bhattacharya-1991}. This contribution increases as $P_{\rm orb}$ decreases, further lowering the $P_{\rm orb}$ value and hence lowering the critical accretion rate (Eqs.~\ref{critical_1} and \ref{critical_2}). This is an additional reason for the relatively longer-lasting persistent phases of short-orbital-period NS LMXBs.

However, there are also aspects that differ between the two critical accretion rate models.
For example, the persistent phase is typically much longer for the $\dot M_{\rm av, crit1}$ model (Eq.~\ref{critical_1}) than for the $\dot M_{\rm av, crit2}$ model (Eq.~\ref{critical_2}).
In fact, in a few cases, the NS LMXB remains transient throughout its evolution for the latter model.
Considering all our computed evolutionary curves (Fig.~\ref{fig: Orbital_Donor Mass}), we find that, for the $\dot M_{\rm av, crit1}$ model (Eq.~\ref{critical_1}), overall, the NS LMXBs are transient during $\sim 88$\% of the total time, while this fraction is $\sim 98$\% for the critical rate, $\dot M_{\rm av, crit2}$ (Eq.~\ref{critical_2}).
This is an effect of different $P_{\rm orb}$, $M_1$ and $M_2$ dependencies of the critical accretion rate for the two models (Eqs.~\ref{critical_1} and \ref{critical_2}; the former does not depend on $M_2$).
Note that this should have implications for observed transient and persistent sources. 
Moreover, for a given set of parameter values, a source could be transient for one critical accretion rate model but persistent for another.
This shows that our computations provide a new way to discriminate among various models of critical accretion rate.

In Fig.~\ref{fig: Orbital_Donor Mass}, we also display observed NS LMXBs with known $M_2$ and $P_{\rm orb}$ (see Table~\ref{Table1: Transient LMXB}).
This is to check whether the computed long-term evolution of NS LMXBs is consistent with the properties of known sources. 
To be more specific, the aim is to check if the known persistent sources are overall attached to the persistent portions  (shown in red) of the computed curves of Fig.~\ref{fig: Orbital_Donor Mass}. The same applies to the known transient sources and the transient portions (shown in green) of the computed curves.
We find a partial consistency, as a fraction of the known transient sources is consistent with the green portion, and a fraction of the known persistent sources is consistent with the red portion (Fig.~\ref{fig: Orbital_Donor Mass}). 
Since the fractional duration of the green portion is less for panel (a) than that for panel (b) of Fig.~\ref{fig: Orbital_Donor Mass}, the $\dot M_{\rm av, crit1}$ model (Eq.~\ref{critical_1}) is better consistent with the known persistent sources, and the $\dot M_{\rm av, crit2}$ model (Eq.~\ref{critical_2}) is better consistent with the known transient sources, overall.

Despite only partial consistency, the following aspect is observed. 
As mentioned above, our computations show that the persistent phase lasts longer in sources with lower $P_{\rm orb}$ values (Fig.~\ref{fig: Orbital_Donor Mass}).
In order to check if this is consistent with the known NS LMXB distribution, we use 29 persistent sources and 50 transient sources with known $P_{\rm orb}$ values \citep{2020ApJS..249...32G, 2023AandA...675A.199A}, and perform a Kolmogorov–Smirnov (KS) test. 
We find that the known persistent source distribution is somewhat shifted towards lower $P_{\rm orb}$ values relative to the known transient source distribution, and the two distributions are separate with a KS statistic value (p-value) of 0.27 (0.11).
While this is a separation of modest significance, it indicates an overall consistency with our computational results. 
Moreover, the range of X-ray luminosities, estimated from persistent-phase accretion rates obtained from our evolutionary computations, is consistent with the observed X-ray luminosities of persistent NS LMXBs.

\section{Discussion and Conclusion} \label{Sec: Discussion and Conclusion}

In this work, we have probed the transient and persistent phases of NS LMXBs using long-term evolution computations. 
As explained in section~\ref{Sec: Intro}, such a study, which we present here for the first time to the best of our knowledge, is crucial for gaining a deeper, broader understanding of these sources and is important for several reasons. 
For example, the identification of persistent and transient phases of NS LMXBs is essential to compute the evolution of the NS properties, e.g., its spin frequency, $\nu$ \citep{2021MNRAS.502L..45B}. Note that the $\nu$ evolution also depends on the disk-magnetosphere interaction, continuous gravitational wave emission due to the mass quadrupole moment of the spinning NS, etc.
Therefore, our study is crucial for probing fundamental NS physics, including the spin-induced continuous gravitational waves, using the observed parameters of RMSPs and NS LMXBs \citep[see section~\ref{Sec: Intro}; ][]{Bhattacharyya2010,BhattacharyyaJ1023,Bhattacharyya2017AQX-1}.
Besides, we find that continuous gravitational radiation due to the orbital motion can become high and may cause a substantial increase in $\dot M_{\rm av}$ in certain phases of the LMXB state (section~\ref{sec: Results}).
Such a high $\dot M_{\rm av}$ can make the source persistent, depending on the specific thermal-viscous instability model and other parameters.

While specific models of accretion, e.g., the ones based on thermal-viscous instability, may explain why an NS LMXB is transient rather than persistent, they do not provide crucial, broad perspectives, such as why there are more transient sources than persistent
ones. 
To the best of our knowledge, there are 84 transient and 43 persistent sources currently known \citep{2023AandA...675A.199A, 2020ApJS..249...32G}.
Moreover, while we expect to have already discovered most bright persistent Galactic NS LMXBs in X-rays, there can be many still unknown transient Galactic NS LMXBs, because the latter sources are detected, or at least identified, only when they undergo outbursts.
This suggests that the number of transient NS LMXBs is much larger than that of persistent ones in the Galaxy.
Our computation of evolution naturally explains this, probably for the first time (see Fig.~\ref{fig: Orbital_Donor Mass}). 
This, irrespective of the selected thermal-viscous instability model, results primarily from how the accretion rate and the orbital period evolve, and mainly from the fact that typically $\dot{M}_{\rm av}$ is initially high and then has a relatively low value for most of the LMXB state duration (see section~\ref{sec: Results}).

Apart from this general explanation, our computations shed light on how various parameters, such as the initial values of $P_{\rm orb}$ and $M_2$, affect the evolution of NS LMXBs, periods of their persistent and transient phases, and how these phases are distributed over the entire LMXB state. 
For example, lower $P_{\rm orb}$ values typically cause longer periods of persistent accretion (see section~\ref{sec: Results}). 
Such dependencies of persistent and transient phases on the nature of binary systems and companion stars could provide new insight into what kinds of sources are transients, what kinds of sources are persistents, and why one observes specific distributions of transients and persistents, for example, in systems like globular clusters, the Galactic plane, etc. 
This can also have implications for transient detection.

We also find  (section~\ref{sec: Results} and Fig.~\ref{fig: Orbital_Donor Mass}) that the periods of persistent and transient accretion sensitively depend on the expression of critical accretion rate $\dot{M}_{\rm av,crit} \propto M_1^l M_2^m P_{\rm orb}^n$, which is a general form of Eqs.~\ref{critical_1} and \ref{critical_2}. 
Thus, our computational results, when compared with the properties of many observed NS LMXBs, provide a new potential way to constrain $l$, $m$, and $n$, and hence the proposed thermal-viscous instability models.
Interestingly, we find that, depending on the instability model and parameter values, a source could remain transient throughout its LMXB state.
As mentioned in section~\ref{Sec: Intro}, a better understanding of transient accretion induced by thermal-viscous instability is essential to probe accretion processes in the strong-gravity regime and estimate related parameter values over a large $\dot{M}$ range.

Finally, from comparing our evolutionary computations with the observed NS LMXB properties, we find the following (see section~\ref{sec: Results} and Fig.~\ref{fig: Orbital_Donor Mass}).
(1) Similar to our computations, the observed NS LMXBs also indicate a greater fraction of persistents with lower orbital periods.
(2) In the $P_{\rm orb}-M_2$ space, the observed NS LMXBs are partially consistent with our computations.
While this partial matching indicates the overall qualitative correctness of thermal-viscous instability models, the lack of full consistency also suggests a need to improve them.
(3) The lack of full consistency could also be partly because Eqs.~\ref{critical_1} and \ref{critical_2} may not be entirely valid for long outburst sources or white dwarf donors \citep[][and references therein]{Bhattacharya:2025qps}.

\section*{Acknowledgements}
DT acknowledges the financial support from the DST-Inspire Scholarship for Higher Education and the role of the MESA community forum in resolving MESA-related queries.  PO acknowledges support by the Polish National Science Center under the grant 2023/50/A/ST9/00527
DT and PO acknowledge Abhijnan Kar for helpful discussions during the project.

\bibliographystyle{aasjournalv7}
\bibliography{paper}

\end{document}